\documentclass[sigplan]{acmart}

\usepackage{natbib}
\usepackage{pgfplots}
\usepackage{tikz}
\usepackage{subcaption}
\usepackage{enumitem}

\pgfplotsset{compat=1.18}
\usepgfplotslibrary{fillbetween}
\usetikzlibrary{positioning,arrows.meta,shapes.geometric,fit,calc}

\setcopyright{none}
\acmDOI{}
\acmISBN{}
\acmConference[SOSP '17]{ACM SIGOPS 31st Symposium on Operating Systems Principles}{2017}{Shanghai, China}
\acmYear{2017}

\title{The Missing Memory Hierarchy: Demand Paging for LLM Context Windows}

\author{Tony Mason}
\affiliation{%
  \institution{University of British Columbia}
  \country{Canada}
}
\affiliation{%
  \institution{Georgia Institute of Technology}
  \country{USA}
}
\email{fsgeek@cs.ubc.ca}
\email{fsgeek@gatech.edu}
\email{fsgeek@wamason.com}



\begin{document}

\begin{abstract}
The context window of a large language model is not memory. It is
L1~cache---a small, fast, expensive resource that the field treats
as the entire memory system. There is no L2, no virtual memory, no
paging. Every tool definition, every system prompt, and every stale
tool result occupies context for the lifetime of the session. The
result is measurable: across 857~production sessions and
4.45~billion effective input tokens, 21.8\% is structural waste.

We present \textsc{Pichay}, a demand paging system for LLM context
windows. Implemented as a transparent proxy between client and
inference API, \textsc{Pichay} interposes on the message stream to
evict stale content, detect page faults when the model re-requests
evicted material, and pin working-set pages identified by fault
history. In offline replay across 1.4~million simulated evictions,
the fault rate is 0.0254\%. In live production deployment over
681~turns, the system reduces context consumption by up to 93\%
(5{,}038KB $\to$ 339KB); under extreme sustained pressure, the
system remains operational but exhibits the expected thrashing
pathology, with repeated fault-in of evicted content.

The key observation is that the problems the field faces---context
limits, attention degradation, cost scaling, lost state across
sessions---are virtual memory problems wearing different clothes.
The solutions exist: working set
theory~\citep{denning1968working}, demand paging, fault-driven
replacement policies, and memory hierarchies with multiple
eviction-managed levels. We describe the architecture of a full
memory hierarchy for LLM systems (L1~through persistent storage),
report on the first three levels deployed in production use
(L1~eviction, L2~fault-driven pinning, L3~model-initiated
conversation compaction), and identify cross-session memory as the
remaining frontier.
\end{abstract}

\maketitle

\section{Introduction}
\label{sec:intro}

In 1961, the Atlas computer introduced demand paging: programs could
address more memory than physically existed, and the hardware would
transparently load pages from backing store as
needed~\citep{kilburn1962one}. Before Atlas, programmers manually
managed overlays---explicitly swapping code segments in and out of a
fixed address space. The transition from overlays to virtual memory
is one of the foundational advances in systems architecture.

Large language model context windows are in the overlay era. Every
agentic AI tool---Claude Code, Cursor, GitHub Copilot,
Windsurf---manually assembles context on every API call: tool
definitions, system prompts, and the complete message history. There
is no eviction policy, no demand loading, no working set estimation.
The default behavior is accumulation until the context limit forces
a crisis.

This paper makes two contributions. First, we provide the empirical
evidence that context management is a memory management problem. We
instrument 857~production sessions (54{,}170 API calls,
4.45~billion effective input tokens) and find that 21.8\% of tokens
are structural waste from three sources: unused tool schemas
(11.0\%), duplicated content (2.2\%), and stale tool results (8.7\%)
reprocessed at a median amplification of 84.4$\times$. These are
not implementation bugs. They are the inevitable consequence of
managing a finite resource (the context window) without the
abstractions that operating systems developed for the analogous
problem fifty years ago.

Second, we build the system. \textsc{Pichay} is a demand paging
system for LLM context windows, implemented as a transparent proxy
between client and inference API. It distinguishes garbage collection
(removing ephemeral tool output that cannot be re-requested) from
paging (evicting addressable content that can be faulted back in).
It detects page faults when the model re-requests evicted content.
It pins working-set pages identified by fault history. And the
eviction summaries it generates---designed as space-saving
markers---function as retrieval handles that models understand
without instruction: ``\texttt{[Paged out: Read file.py (8,192
bytes). Re-read if needed.]}''

In production deployment, the system extends a session from 7\%
free context to 43\% free---the difference between imminent context
death and substantial remaining capacity. Over a 681-turn session,
it sustains operation at 97\% eviction rate, though this extreme
case reveals the thrashing pathology well-known from OS
research~\citep{denning1968working}: when the working set exceeds
the resident set, the system spends more effort on faulting than on
useful work.

The deeper observation is that the context window is not memory. It
is L1~cache. And nobody is building the rest of the hierarchy.
The field's response to context limits is to make L1 bigger---1M
token windows, 10M token windows. This is the equivalent of building
machines with more physical RAM instead of inventing virtual memory.
It works, but it doesn't scale, and it misses the architectural
insight: a managed hierarchy of small-fast-expensive and
large-slow-cheap storage, connected by eviction policies and fault
mechanisms, outperforms any single level no matter how large.

We describe the full hierarchy: L1~(the generation window), L2~(the
working set, demand-paged and pinned), L3~(session history,
compressed with declared losses), and L4~(cross-session persistent
memory, indexed and retrievable). The first two levels are deployed
and evaluated. L3~is implemented---the model can compress conversation
turns into outcome summaries via cooperative cleanup tags---but not
yet evaluated at scale. L4~is designed and its mechanisms identified.

The contributions:
\begin{enumerate}
  \item A large-scale empirical characterization of context
    window utilization in a production agentic AI corpus, establishing that
    21.8\% of tokens are structural waste with a measured taxonomy.
  \item \textsc{Pichay}, a demand paging system for LLM context
    windows with empirically measured fault rate (0.0254\% offline,
    deployed in production).
  \item Fault-driven pinning: a page replacement policy that
    learns from its own mistakes---one fault pins a page for the
    session, reducing repeated faults in steady-state workloads.
  \item Cooperative memory management via two side channels
    (phantom tools and cleanup tags): mechanisms enabling the model
    to voluntarily release cold pages, request cached faults, and
    compress conversation history---a new point in the design space
    not available in hardware memory hierarchies, where applications
    are non-cooperative.
  \item The architectural observation that LLM context management
    maps structurally (not merely metaphorically) to virtual memory,
    and that the full OS memory hierarchy---from cache replacement
    policies to working set theory---applies directly.
  \item Design of a four-level memory hierarchy for LLM systems,
    with the first three levels deployed (L1~eviction, L2~pinning,
    L3~cooperative compaction) and the first two evaluated.
\end{enumerate}


\section{Background}
\label{sec:background}

\subsection{Agentic AI Architecture}

Modern AI coding assistants operate as agentic systems: the model
receives a set of tool definitions (JSON schemas describing available
operations), a system prompt (identity, instructions, behavioral
constraints), and a growing message history (alternating user and
assistant turns including tool invocations and their results).

On each API call, the client application assembles the full context:
system prompt, tool definitions, and complete message history. The
inference provider tokenizes this context, computes attention over
all tokens, and generates a response. The response is appended to
the message history, and the cycle repeats.

The critical property: \emph{context is reassembled from scratch on
every API call.} There is no persistent state at the inference layer
between calls. The entire conversation---including tool definitions
that haven't changed and results that will never be referenced
again---is sent, tokenized, and attended to on every turn.

\subsection{Context Windows as Physical Memory}

The analogy between context windows and physical memory is structural,
not merely metaphorical (Table~\ref{tab:vm-analogy}).

\begin{table}[ht]
\centering
\caption{Virtual memory analogy for context window management.}
\label{tab:vm-analogy}
\small
\begin{tabular}{@{}ll@{}}
\toprule
\textbf{OS Concept} & \textbf{Context Window Equivalent} \\
\midrule
Physical memory & Context window (200K tokens) \\
Virtual memory & Persistent state (disk, databases) \\
Page table & Retrieval handles (UUIDs, anchors) \\
Page fault & Model re-requests evicted content \\
MMU & Proxy layer between client and API \\
Working set & Currently relevant context subset \\
Demand paging & Load tool definitions on first use \\
Thrashing & Evict/fault cycle exceeding useful work \\
Pinning & Fault history prevents re-eviction \\
\bottomrule
\end{tabular}
\end{table}

Early computer systems managed limited physical memory through
manual overlays: programmers explicitly swapped code segments in and
out of a fixed address space. The transition to virtual
memory---Atlas~\citep{kilburn1962one}, Multics, then
Unix---automated this management through hardware page tables,
demand paging, and working-set-based eviction
policies~\citep{denning1968working,denning1970virtual}.

Context windows are in the overlay era. Applications manually
manage what fits, with no system-level support for eviction,
demand loading, or working set estimation.

\subsection{The Memory Hierarchy}
\label{sec:hierarchy}

The field's response to context pressure is to increase context
window size---from 4K to 32K to 128K to 1M tokens and beyond. This
is the equivalent of building machines with more physical RAM
instead of inventing virtual memory. It works but does not scale:
attention cost is quadratic in context length, and the $O(n^2)$ cost
explosion in long sessions means that even million-token windows
fill and degrade~\citep{zeyliger2026quadratic}.

The alternative is architectural: a managed hierarchy of levels,
each larger, slower, and cheaper than the one above it, connected by
eviction policies and fault mechanisms (Table~\ref{tab:hierarchy}).

\begin{table}[ht]
\centering
\caption{Memory hierarchy for LLM systems.}
\label{tab:hierarchy}
\small
\begin{tabular}{@{}llll@{}}
\toprule
\textbf{Level} & \textbf{Contents} & \textbf{Eviction} & \textbf{Fault} \\
\midrule
L1 & Generation window & --- & --- \\
L2 & Working set (pinned) & Pressure-based & Re-read \\
L3 & Session history & Age-based & Summary expansion \\
L4 & Cross-session memory & LRU / relevance & Retrieval query \\
Storage & Full corpus & None & Search + ingest \\
\bottomrule
\end{tabular}
\end{table}

\textbf{L1} is the active generation window---the tokens the model
attends to on the current API call. Small, fast, expensive per
token. This is what everyone calls ``the context window.''

\textbf{L2} is the working set: content the model is actively using
but that need not occupy L1 on every turn. Managed by fault-driven
pinning (Section~\ref{sec:pinning}). A page evicted from L1 that is
immediately re-requested is promoted to L2 and kept resident until
context pressure forces demotion.

\textbf{L3} is session history: earlier conversation turns,
completed tool interactions, superseded plans. Compressed with
declared losses into structured summaries. Faultable by expanding
the summary back to original content from the client's backing
store.

\textbf{L4} is cross-session persistent memory: authored
compressions (tensors), activity records, semantic indices. Survives
session death. Retrieved by graph traversal or similarity search.

\textbf{Storage} is the full corpus: every conversation transcript,
every tool output, every document. Archived, indexed, never resident
unless demanded.

Content migrates between levels based on access patterns. The hot
path stays in L1. Working-set content is demand-paged between L1 and
L2. Cold history compresses into L3. Cross-session knowledge
persists in L4. The context limit does not disappear---it becomes
the L1~cache size. The total addressable memory is unbounded. System
performance is determined by hit rates at each level, not by L1
capacity alone.

This paper implements and evaluates L1~management (eviction and
garbage collection) and L2~management (fault-driven pinning).
L3~(rolling conversation compaction) is implemented via cooperative
cleanup tags (Section~\ref{sec:l3}) but not yet evaluated at scale.
We describe the interface to L4~(persistent memory stores).

\subsection{Related Work}
\label{sec:related}

\paragraph{Context compression.}
LLMLingua~\citep{jiang2024longllmlingua} and related approaches compress
prompts by removing tokens predicted to be low-information. These
are lossy, model-dependent operations applied to content. Our
interventions are structural: we remove entire categories of waste
(unused tool schemas, duplicate content, dead results) without lossy
compression of the remaining content.

\paragraph{Agent context pruning.}
SWE-Pruner~\citep{wang2026swepruner} trains a 0.6B-parameter neural
skimmer for task-aware context pruning in SWE-bench agents, achieving
23--54\% token reduction. \citet{lindenbauer2025complexity} show that
simple observation masking halves agent cost while matching
LLM-summarization solve rates. ACON~\citep{kang2025acon} provides a
unified framework for history and observation compression, reducing
peak usage by 26--54\%.

These approaches frame the problem as prompt optimization or
compression. The key difference: they optimize for benchmark scores
on fixed task distributions. We measure production waste profiles
and implement eviction policies with an empirically measured fault rate. Their
framing is ``make the prompt smaller.'' Ours is ``manage the working
set.'' Working-set management composes with other interventions and
provides measurable fault rates.

\paragraph{Quadratic cost analyses.}
Recent work has quantified the $O(n^2)$ cost explosion in long
agentic sessions where cache reads dominate~\citep{zeyliger2026quadratic}.
Each API call reprocesses the full context; as sessions grow, the
cumulative cost grows quadratically with session length. Our
amplification factor measurement (Section~\ref{sec:amplification})
confirms this at production scale and our interventions directly
address it.

\paragraph{Prompt caching.}
Anthropic and OpenAI offer prompt caching that avoids recomputation
of static prefixes. In our corpus, 93.5\% of input tokens are cache
reads---caching is working. But cached tokens still occupy the
context window and require attention computation for every output
token generated. Caching reduces the input processing cost; it does
not reduce the attention cost or the memory pressure.

\paragraph{Context engineering.}
The emerging discipline of context
engineering~\citep{deepset2025context,willison2025context,mei2025survey}
recognizes that managing what enters context is as important as the
model itself. \citet{mei2025survey} survey the field comprehensively,
covering retrieval, processing, and management. Anthropic's own
engineering team describes practical
patterns~\citep{rajasekaran2025context}. Our work provides a specific
mechanism (proxy-layer interposition with eviction policies) and the
empirical production data that the field currently lacks.

\paragraph{Prompt-level memory management.}
MemGPT~\citep{packer2023memgpt} pioneered the OS virtual memory
analogy: a FIFO message queue with recursive summarization at
capacity, plus archival/recall databases accessible via tool calls.
The model can save information but cannot explicitly release it;
eviction is capacity-triggered (flush at 100\%), not cost-driven.
Contextual Memory Virtualisation~\citep{cmv2026} models session
history as a DAG with structurally lossless trimming that strips
mechanical bloat (raw tool outputs, base64, metadata) while
preserving conversational content verbatim---achieving up to 86\%
reduction for tool-heavy sessions. Focus~\citep{verma2026focus}
gives the agent autonomy over compression: the LLM consolidates
learnings into a persistent knowledge block and prunes raw history,
inspired by slime mold exploration. SideQuest~\citep{sidequest2026}
teaches the model to evict stale tool outputs from the KV cache via
a fine-tuned parallel reasoning thread, achieving 56--65\% peak
memory reduction---but eviction is irreversible and requires
inference engine modifications.

These systems address the architectural problem (MemGPT's VM
analogy), the structural problem (CMV's lossless trimming), the
autonomy problem (Focus's model-directed compression), or the
inference problem (SideQuest's KV cache eviction). None derives an
economic model where prompt-token retention cost scales quadratically
while fault cost scales linearly. None provides cooperative
recall/release primitives informed by that cost model. None reports
empirical fault rates on production sessions.

\paragraph{Positioning.}
Prior work addresses either the cost problem (quadratic analyses) or
the content problem (compression, pruning, RAG) but not the
structural problem: context windows are unmanaged physical memory.
Our contribution is the systems abstraction---working set, eviction
policy, fault rate, amplification factor---and the empirical evidence
that it works on production data.


\section{System Design}
\label{sec:system}

\textsc{Pichay} is a transparent HTTP proxy that interposes between
an agentic AI client and the inference API. On each request, it
receives the full message array assembled by the client, applies
context management policies, and forwards the modified request to
the inference provider. The client is unaware of the interposition.

\subsection{Architecture}

The proxy operates at the Messages API level. It receives JSON
requests containing the system prompt, tool definitions, and message
history. It returns the inference provider's response unmodified.
The interposition point is the message array: the proxy can inspect,
measure, modify, or replace any message before forwarding.

This design has three properties: (1)~it requires no changes to the
client, the model, or the inference API; (2)~the client retains the
full, unmodified conversation history, providing a backing store
from which evicted content can be faulted back in; and (3)~the proxy
sees the complete request on every turn, enabling stateful
management without persistent connections.

\subsection{Garbage Collection vs.\ Paging}

Not all tool results are equal. We distinguish two categories:

\textbf{Garbage collection} applies to ephemeral tool outputs---Bash
command results, search outputs, directory listings---that have no
stable identity. Once consumed, they cannot be meaningfully
re-requested. Removing them is reclamation of dead content.

\textbf{Paging} applies to addressable content---file reads, plan
documents, specifications---that have stable identity (a file path).
Removing these creates a fault risk: the model may need the content
again and must re-read the file.

The distinction matters for fault rate calculation. Only paging
(Read evictions) can produce faults. Garbage collection cannot.
Conflating them inflates the eviction denominator and deflates the
apparent fault rate.

\subsection{Eviction Policy}

The current policy is FIFO by user-turn age: tool results older than
$\tau$~user turns and larger than $s_{\min}$~bytes are candidates
for eviction ($\tau = 4$, $s_{\min} = 500$ in all experiments). This
is deliberately simple. The research question is how far a minimal
policy extends before requiring sophistication.

Evicted content is replaced with a retrieval handle:

\begin{quote}
\small\texttt{[Paged out: Read /path/to/file.py (8,192 bytes,
187 lines). Re-read the file if you need its content.]}
\end{quote}

The handle serves three functions: (1)~it identifies what was
removed; (2)~it states how to recover the content; and (3)~it
occupies substantially less context than the original.

\subsection{Page Fault Detection}

When the model issues a tool call that matches an evicted entry (same
tool name and arguments), the proxy detects a \emph{page fault}: the
model is requesting content it previously had but lost to eviction.
The evicted entry is indexed by a key derived from the tool name and
input arguments (e.g., file path for Read operations).

Page faults are observable events. They measure the cost of eviction:
each fault is a wasted tool call round-trip that would not have
occurred if the content had remained resident. At inference-time
pricing, faults have direct monetary cost.

\subsection{Fault-Driven Pinning}
\label{sec:pinning}

The simplest upgrade to FIFO: if evicting a page caused a fault,
don't evict it again. One fault per file, permanently pinned for the
session. The algorithm:

\begin{enumerate}[nosep]
  \item On eviction of addressable content: record the file path and
    content hash.
  \item On page fault: record the evicted content's hash in a fault
    history table.
  \item On next eviction attempt for the same path: if the current
    content hash matches the fault history entry, pin the page---skip
    eviction permanently.
  \item On new read of a pinned path with different content (file was
    edited): unpin. The old version is stale; the new version starts
    a fresh fault cycle.
\end{enumerate}

The content hash comparison prevents false pins: if a file changed
between eviction and re-read, the eviction was correct (stale data
was removed). Pinning only applies when the model demonstrably
needed exactly what was taken away. The unpin-on-edit rule handles
the edit/review cycle common in coding sessions.

\subsection{Retrieval Handles as Anchors}

The eviction summary was designed as a space-saving marker. In
practice, it functions as a late-binding retrieval handle---an anchor
that stores minimal metadata and resolves to full content on demand.

The handle resolves to \emph{current} content, not the original
evicted content. A file edited since eviction materializes at its
new state when faulted in. This temporal property is a feature: the
model always gets the latest version, and stale cached content is
impossible.

Behavioral evidence: when a fresh model instance resumed a session
containing paged-out content, it unprompted stated: ``Let me re-read
the files I need since some were paged out.'' The model recognized
the handles, understood what was missing, and chose to fault content
in before acting. The handle's format carries its own semantics---no
instruction was needed.

\subsection{Cooperative Memory Management}
\label{sec:cooperative}

In OS virtual memory, the application is non-cooperative. It never
voluntarily releases pages. The OS must infer the working set from
access patterns because the CPU will not say what it needs next.
Decades of replacement algorithms---FIFO, LRU, Clock, Working
Set---exist because of this fundamental asymmetry.

LLMs break this assumption. The model has \emph{incentive} to
cooperate with the memory manager: cleaner context means better
attention, better output quality, and longer session life. The
processor wants to help manage its own cache. This changes the
design space.

\textsc{Pichay} exploits this through \emph{phantom tools}---tool
definitions injected by the proxy that the framework never sees.
The model can call them; the proxy intercepts the calls from the
streaming response before the framework receives them, handles
them internally, and injects coherent tool results on the next
turn. The framework is unaware of the side channel.

Two phantom tools are defined:

\begin{description}[nosep]
  \item[\texttt{memory\_release(paths)}] The model signals it no
    longer needs specific files. The proxy marks them for immediate
    eviction, bypassing the age threshold. This is the reference
    bit---provided voluntarily by the processor.
  \item[\texttt{memory\_fault(paths)}] The model requests evicted
    content restored from the proxy's backing store. The proxy
    resolves the fault from its eviction cache without a file system
    round trip---faster and cheaper than a real Read tool call.
\end{description}

The second cooperative channel is \emph{cleanup tags}---structured
directives embedded in the model's output text, parsed by the proxy
before forwarding to the framework. Where phantom tools are
proxy-to-model (the proxy offers capabilities), cleanup tags are
model-to-proxy (the model initiates management). Four operations are
defined:

\begin{description}[nosep]
  \item[\texttt{drop: block:ID}] Immediately evict a specific block.
  \item[\texttt{summarize: block:ID "text"}] Replace a block with a
    compact summary---lossy compression authored by the model, not
    the system.
  \item[\texttt{anchor: block:ID}] Pin a block against eviction.
  \item[\texttt{collapse: turns N-M "text"}] Compress a range of
    conversation turns into a single synthetic summary block,
    removing all intermediate content.
\end{description}

The collapse operation is particularly significant: it enables L3
compaction (Section~\ref{sec:l3}) initiated by the model itself,
based on its understanding of which dialogue was scaffolding and
which produced durable outcomes.

The key insight: in hardware memory hierarchies, the application is
adversarial or indifferent. The entire replacement algorithm
literature assumes non-cooperation. Cooperative demand paging---where
the processor voluntarily releases cold pages and explicitly
requests faults---is a new point in the design space, enabled by the
fact that LLMs experience degraded output quality under context
pressure and can learn that cooperation improves their own
performance. The two channels (phantom tools and cleanup tags)
provide both directions of the cooperative protocol: the proxy
informs the model of memory state, and the model directs the proxy
to act on its cognitive assessment of what is cold.

\subsection{Graduated Pressure Zones}
\label{sec:pressure}

The eviction policy must respond to context pressure, but a binary
threshold (evict/don't-evict) is too coarse. \textsc{Pichay} defines
four pressure zones based on token consumption:

\begin{description}[nosep]
  \item[Normal] ($< 60\text{K}$ tokens): No intervention. The proxy
    observes and logs.
  \item[Advisory] ($60\text{K}$--$100\text{K}$): The proxy injects
    memory pressure information into the model's context---current
    fill percentage, the five largest resident blocks, and available
    cleanup operations. The model can act cooperatively or ignore the
    advisory. This is the graduated equivalent of the ``low memory''
    notification in desktop operating systems.
  \item[Involuntary] ($100\text{K}$--$120\text{K}$): The proxy begins
    automatic eviction using the FIFO age policy. The model is
    informed but not consulted.
  \item[Aggressive] ($\geq 120\text{K}$): Emergency eviction with
    relaxed thresholds. Context survival takes priority over working
    set preservation.
\end{description}

The advisory zone is the cooperative innovation. Rather than
evicting silently (the OS approach) or crashing at capacity (the
current agentic AI approach), the system provides the model with
enough information to make intelligent cleanup decisions. The
60K threshold was chosen to provide approximately 40K tokens of
runway before involuntary eviction---enough for the model to
complete a coherent thought and emit cleanup tags before losing
agency over the process.

\subsection{L3: Rolling Conversation Compaction}
\label{sec:l3}

The proxy manages tool results (L1/L2). Non-tool content---conversation
history, planning dialogue, orientation reads---is managed by the
collapse operation (Section~\ref{sec:cooperative}). This is L3 in the
memory hierarchy: session history compressed with declared losses into
structured summaries.

The collapse operation \texttt{collapse: turns N-M "summary"} replaces
all blocks in a contiguous turn range with a single synthetic block
containing the model-authored summary. The original content is not
stored---this is lossy compression by design. The summary captures
\emph{outcomes} (what was decided, what was built, what failed) rather
than \emph{process} (which files were read, what intermediate steps
occurred).

Block state persists across session restarts via atomic checkpointing.
On each cleanup tag processing pass, the proxy serializes block metadata
(IDs, content hashes, sizes, turns, roles, status, summaries) to a JSON
checkpoint file using atomic write (tmp file + rename). On session
creation, the checkpoint is loaded and block tracking resumes from the
persisted state. Original content is lazily repopulated as the message
array is reprocessed.

The checkpoint is metadata-only---content is not persisted (the client's
message array is the backing store). This keeps checkpoint files small
(kilobytes, not megabytes) and avoids the consistency hazard of
maintaining two copies of message content.


\section{Measurement}
\label{sec:methodology}

\subsection{Corpus}
\label{sec:corpus}

We instrument Claude Code, Anthropic's AI coding assistant, across
a single power user's sessions over approximately four months
(November 2025 through March 2026). The corpus comprises 857
sessions across 15 software projects, drawn from two sources:
a local WSL environment and an Ubuntu VM (migrated from a prior
machine).

Sessions are classified by type:
\begin{itemize}[nosep]
  \item \textbf{Main sessions} (59): Human-facing conversations
    where the user interacts directly with the assistant.
  \item \textbf{Subagent sessions} (567): Delegated tasks spawned
    by the main session to handle subtasks in parallel.
  \item \textbf{Compact sessions} (154): Context compaction
    continuations triggered when the main session approaches
    its context limit (the client automatically summarizes
    conversation history to stay within the context window).
  \item \textbf{Prompt suggestion sessions} (21): Short sessions
    for generating prompt suggestions.
\end{itemize}

The corpus represents 54{,}170 API calls and 4.45~billion effective
input tokens (input tokens + cache creation + cache read tokens).
Token usage is extracted from the API response metadata embedded in
session transcripts.

Because the live corpora continue to evolve, the paper reports a
frozen cohort (857 sessions) rather than a moving live count. We
provide a recount script (\texttt{tools/corpus\_counts.py}) to make
the counting boundary and dedup assumptions explicit.
Unless stated otherwise, all corpus tables and figures in this draft
use the frozen cohort with a fixed counting protocol
(\texttt{dedup\_by\_content}, \texttt{min\_size=10000}) so results are
stable across revisions.

\paragraph{Bias declaration.}
This is a single-user corpus from a power user who regularly
exhausts context windows across complex, multi-file engineering
tasks. We argue this is representative of the class where context
management matters most, and note that Pareto-distributed usage
means this class likely dominates total fleet token consumption.
Multi-user validation is identified as future work.

\subsection{Instruments}

We develop three instruments at different layers of the stack:

\paragraph{Probe} (\texttt{probe.py}): A streaming JSONL analyzer
that reads Claude Code's raw session transcripts. Classifies records
by type (user, assistant, tool\_result, progress), measures content
sizes, tracks tool usage, and computes per-session metrics including
amplification factor and tool overhead ratio. Requires no API calls;
operates on existing session files.

\paragraph{Proxy} (\texttt{proxy.py}): A transparent HTTP proxy
deployed between Claude Code and the Anthropic API. Intercepts
every request, capturing system prompts, tool definitions, message
arrays, and token usage. Optionally applies interventions (eviction,
trimming) before forwarding. Logs all decisions to JSONL for offline
analysis. This is the ``MMU'' in our analogy.

\paragraph{Experiment runner} (\texttt{pichay}): An experiment
framework that orchestrates paired runs under different treatment
conditions (baseline, trimmed, compact+trim), captures all
artifacts (proxy logs, session data, git state, test results),
and computes comparative metrics. The framework is self-bootstrapping:
it was built by Claude Code running through its own proxy, with
each round's logs feeding the next round's analysis.

\subsection{Treatment Conditions}

We define three treatment conditions applied at the proxy layer:

\begin{enumerate}[nosep]
  \item \textbf{Baseline}: Proxy in observation mode. Logs all
    traffic without mutation.
  \item \textbf{Trimmed}: Tool definition stubbing and skill
    deduplication applied to each request before forwarding.
  \item \textbf{Compact+Trim}: Stale result eviction (paging)
    combined with trimming.
\end{enumerate}

All treatments use the same model (Claude Opus), same temperature,
and same task prompts. The proxy's interposition is transparent to
both the client and the API---neither is aware of the mutations.


\section{Results}
\label{sec:results}

The analysis proceeds in two phases. Phase~1 characterizes broad
conversational patterns---tool overhead ratios, amplification factors,
and token accounting---across the full 857-session corpus using session
transcript analysis. Phase~2 performs detailed API-level waste
decomposition on a smaller proxy-captured sample (5~sessions,
99~API calls) where we have full request payloads. The corpus-scale
projection (Section~\ref{sec:corpus-projection}) then bridges the two
phases, applying measured conversion constants from Phase~2 to the
full corpus to estimate fleet-wide addressable waste.
The conversion constant used for projection (bytes-to-token ratio)
is estimated from a broader proxy sample of 139~API calls.

\subsection{Phase 1: Conversation-Level Waste}
\label{sec:phase1}

\paragraph{Tool overhead.}
Across the full corpus, 79.4\% of conversation bytes are tool
results. Assistant text accounts for 12.7\% and user text for
7.9\%. Two independent measurements converge on this ratio
(78.2\% and 79.4\%), suggesting it is a stable property of the
workload, not an artifact of sampling.

\begin{figure}[ht]
\centering
\begin{tikzpicture}
\begin{axis}[
    xbar,
    width=0.85\columnwidth,
    height=8cm,
    xlabel={Adoption rate (\% of sessions)},
    symbolic y coords={
        EnterWorktree, NotebookEdit, TodoWrite, Skill,
        Agent, TaskStop, TaskOutput, AskUser,
        EnterPlan, ExitPlan, WebSearch, WebFetch,
        Edit, Write, Grep, Glob, Bash, Read
    },
    ytick=data,
    yticklabel style={font=\small},
    xmin=0, xmax=80,
    bar width=6pt,
    nodes near coords={\pgfmathprintnumber\pgfplotspointmeta\%},
    nodes near coords style={font=\tiny, anchor=west},
    every axis plot/.append style={fill=blue!40},
]
\addplot coordinates {
    (0.0,EnterWorktree) (0.0,NotebookEdit) (0.0,TodoWrite)
    (0.1,Skill) (0.9,Agent) (0.9,TaskStop)
    (1.9,AskUser) (3.1,TaskOutput)
    (3.4,EnterPlan) (3.9,ExitPlan) (3.7,WebSearch)
    (3.9,WebFetch) (16.0,Edit) (23.6,Write)
    (25.7,Grep) (37.5,Glob) (63.0,Bash) (72.7,Read)
};
\end{axis}
\end{tikzpicture}
\caption{Tool adoption rate across 801 active sessions. The median
  session uses 3 of 18 available tools. Seven tools see zero or
  near-zero adoption, yet their full schemas (collectively
  $\sim$24{,}500 bytes) are sent on every API call.}
\label{fig:tool-adoption}
\end{figure}
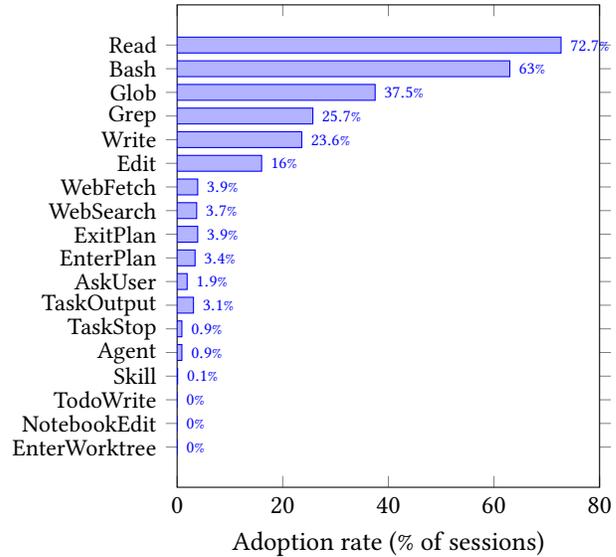

\paragraph{Tool type concentration.}
Tool usage is extremely concentrated (Figure~\ref{fig:tool-adoption}).
\texttt{Read} accounts for 75\% of all tool output bytes (9{,}393
calls, mean 7{,}935 bytes per result). \texttt{Bash} accounts for
13.3\% (10{,}090 calls). All other tools contribute less than 5\%
each. This concentration has a direct implication: file content
read by the \texttt{Read} tool is the dominant source of
context bloat.

\paragraph{Amplification factor.}
\label{sec:amplification}
Each tool result persists in context from the turn it was generated
until the session ends (or the client's internal compaction evicts
it). We define the \emph{amplification factor} as:
\[
  A = \frac{\sum_{r \in R} \text{size}(r) \times \text{turns\_survived}(r)}
           {\sum_{r \in R} \text{size}(r)}
\]
where $R$ is the set of tool results and turns\_survived is the
number of subsequent turns the result remains in context. This
measures how many times, on average, each byte of tool output is
reprocessed.

For main sessions: median $A = 84.4\times$, P75 $= 217.9\times$,
P90 $= 570.8\times$. For subagents: median $A = 12.8\times$
(short-lived, less accumulation). Amplification scales linearly
with session length at a ratio of approximately 0.5, confirming
the quadratic cost structure identified in recent
analyses~\citep{zeyliger2026quadratic}.

\paragraph{Token accounting.}
The corpus consumed 4.45~billion effective input tokens with a
93.5\% cache hit ratio, meaning the vast majority of input
processing is served from cache. However, cached tokens still
occupy the context window and require attention computation for
every output token. The average API call processes 82{,}061
effective input tokens and generates 88 output tokens---a
933:1 input-to-output ratio. Agentic coding is an overwhelmingly
input-bound workload.

\subsection{Phase 2: API-Level Waste Taxonomy}
\label{sec:taxonomy}

Using the proxy, we decompose 5 sessions (99 API calls, 24.4~MB
of request data) into four categories of addressable waste
(Table~\ref{tab:waste-taxonomy}).

\begin{table}[ht]
\centering
\caption{Waste taxonomy. Percentages are of total request bytes
  across 99 API calls.}
\label{tab:waste-taxonomy}
\small
\begin{tabular}{@{}lrrl@{}}
\toprule
\textbf{Category} & \textbf{Bytes} & \textbf{\%} & \textbf{Mechanism} \\
\midrule
Dead tool output & 6{,}468{,}360 & 26.5 & Stale results never re-referenced \\
Tool definition stubs & 4{,}924{,}950 & 20.2 & Schemas for unused tools \\
Static re-send & 2{,}680{,}794 & 11.0 & Unchanged system prompt content \\
Skill triplication & 700{,}582 & 2.9 & Same skill listed $3\times$ \\
\midrule
\textbf{Total addressable} & \textbf{14{,}774{,}686} & \textbf{60.5} & \\
\bottomrule
\end{tabular}
\end{table}

\paragraph{Dead tool output (26.5\%).}
Tool results from early turns persist in context long after any
reference to them. A file read during orientation (Q1 of the
session) survives approximately 90\% of the remaining session.
The content is reprocessed on every subsequent API call, consuming
attention without contributing to the current task.

\paragraph{Tool definition schemas (20.2\%).}
Claude Code sends 18 tool definitions totaling 63{,}088 bytes on
every API call. The median session uses 3 tools. The remaining 15
tool schemas ($\sim$52{,}500 bytes) are sent, tokenized, and
attended to on every call without ever being invoked.

\paragraph{Static system content (11.0\%).}
The system prompt, CLAUDE.md instructions, and memory files are
re-sent identically on every API call. While prompt caching avoids
the forward-pass cost, these tokens still occupy the context window
and participate in attention computation.

\paragraph{Skill triplication (2.9\%).}
The skills list---describing available slash commands---is injected
into messages three times under different prefixes (\texttt{base},
\texttt{example-skills:base}, \texttt{document-skills:base}).
Simple deduplication removes two-thirds of the entries.

\subsection{Interventions}
\label{sec:interventions}

All interventions operate in the proxy layer between the client
application and the inference API. No changes to the model, the
client, or the API are required.

\paragraph{Tool definition stubbing.}
Unused tool definitions are replaced with minimal stubs:
\begin{verbatim}
{"name": "NotebookEdit",
 "description": "...(first line)...",
 "input_schema": {"type":"object","properties":{}}}
\end{verbatim}
Full schemas ($\sim$3{,}505 bytes each) are replaced with stubs
($\sim$80 bytes). On first invocation of a stubbed tool, the full
definition is restored from a stored copy. The intervention is
session-scoped: once a tool is used, its schema remains restored
for the session. Per-request savings: $(18 - k) \times 3{,}425$
bytes, where $k$ is the number of tools used so far.

\paragraph{Content deduplication.}
Skill entries are parsed, grouped by base name, and deduplicated
(keeping the first occurrence). Saves 7{,}453 bytes per request.
Static system prompt components are tracked by content hash across
turns; identical components are logged as candidates for prefix
caching. Currently measurement-only for static components---actual
stripping requires cache-aware API support.

\paragraph{Stale result eviction (paging).}
Tool results older than a threshold (4 user-turns from the end of
the conversation) and larger than a minimum size (500 bytes) are
evicted. Error results are never evicted (the model needs them for
debugging). Evicted content is replaced with a summary:
\begin{quote}
\small
\texttt{[Paged out: Read /path/to/file.py (12{,}450 bytes, 287
  lines). Re-read if needed.]}
\end{quote}
The summary preserves the tool name, key parameter, and original
size ($\sim$200 bytes regardless of original size). If the model
re-invokes the same tool with the same parameters, this constitutes
a \emph{page fault}---the eviction was incorrect and the content
was still needed.

\subsection{Eviction Safety}
\label{sec:eviction}

We validate eviction safety through offline replay: 29
proxy-captured sessions are replayed through the pager, simulating
eviction decisions without making API calls.

\begin{table}[ht]
\centering
\caption{Eviction safety results from offline replay.}
\label{tab:eviction}
\small
\begin{tabular}{@{}lr@{}}
\toprule
Total simulated evictions & 1{,}393{,}000 \\
Page faults detected & 354 \\
\textbf{Fault rate} & \textbf{0.0254\%} \\
Content evicted & 8.49 GB \\
\bottomrule
\end{tabular}
\end{table}

A fault rate of 0.0254\% means the eviction policy is sound:
content older than 4 user-turns is almost never needed again. The
354 faults represent content that was genuinely dead by the age
criterion but happened to match a later request. A fault rate of
zero would indicate over-conservative eviction; some faults are
expected and acceptable.

\subsection{Live Treatment Comparison}
\label{sec:treatment}

We run a standardized task under three treatment conditions
(Table~\ref{tab:treatment}).

\begin{table}[ht]
\centering
\caption{Treatment comparison on a standardized task.}
\label{tab:treatment}
\small
\begin{tabular}{@{}lrrr@{}}
\toprule
\textbf{Metric} & \textbf{Baseline} & \textbf{Trimmed} & \textbf{Compact+Trim} \\
\midrule
API calls & 3 & 4 & 3 \\
Effective input tokens & 114{,}222 & 88{,}421 & 71{,}816 \\
Cache read tokens & 79{,}712 & 38{,}639 & 32{,}228 \\
Task completed correctly & Yes & Yes & Yes \\
\midrule
\textbf{Token reduction} & --- & 22.6\% & \textbf{37.1\%} \\
\bottomrule
\end{tabular}
\end{table}

The compact+trim treatment achieves 37.1\% reduction in effective
input tokens. Cache reads drop 59.6\%. The task completes correctly
under all conditions in this controlled run.

\subsection{Corpus-Scale Projection}
\label{sec:corpus-projection}

We validate the small-sample findings across the full 857-session
corpus using measured constants from the proxy sessions and tool
usage patterns extracted from raw session transcripts.

The bytes-to-token conversion ratio of 4.15 bytes per effective
input token is measured from 139 proxy-captured API calls.

\begin{table}[ht]
\centering
\caption{Corpus-scale token savings projection.}
\label{tab:corpus-tokens}
\small
\begin{tabular}{@{}lrr@{}}
\toprule
\textbf{Intervention} & \textbf{Tokens Saved} & \textbf{\% of Input} \\
\midrule
Tool stub trimming & 487.5M & 11.0\% \\
Skill deduplication & 95.8M & 2.2\% \\
Static re-send & 387.0M & 8.7\% \\
\midrule
\textbf{Total addressable} & \textbf{970.4M} & \textbf{21.8\%} \\
\bottomrule
\end{tabular}
\end{table}

The average saving is 17{,}913 tokens per API call. Across the
corpus, this translates to 85~billion fewer token-token attention
pairs (17{,}913 context tokens $\times$ 88 output tokens $\times$
54{,}170 calls).

\subsection{Production Deployment}
\label{sec:production}

We deploy \textsc{Pichay} in compact mode as the proxy for our own
development workflow---the system is used daily by the authors for
software engineering tasks. This section reports on two
representative sessions that illustrate different operating regimes.

\paragraph{Session A: Steady-state coding.}
A standard coding session using compact mode ($\tau = 4$,
$s_{\min} = 500$):

\begin{table}[ht]
\centering
\caption{Production Session A: steady-state coding.}
\label{tab:session-a}
\small
\begin{tabular}{@{}lr@{}}
\toprule
Context free before compaction & 7\% \\
Context free after compaction & 43\% \\
Total evictions & 15 \\
Garbage collected & 11 \\
Read evictions (pageable) & 4 \\
Page faults & 1 \\
Fault rate (Read only) & 25\% \\
\bottomrule
\end{tabular}
\end{table}

The session recovered 36~percentage points of context capacity---the
difference between imminent context exhaustion and substantial
remaining headroom. Of 15~evictions, 11~were garbage collection
(Bash/Grep/Glob outputs) and 4~were Read evictions. The single
fault was a plan file read early in the session and evicted when it
crossed the age threshold. The plan file is reference material
needed for the entire session---a hot page that FIFO treats as cold.
This is the classic working set failure: the eviction policy measures
age, not access pattern.

\paragraph{Session B: Sustained multi-agent coordination.}
A 681-turn session involving multiple subagent spawns and
cross-project file analysis:

\begin{table}[ht]
\centering
\caption{Production Session B: 681-turn sustained session.}
\label{tab:session-b}
\small
\begin{tabular}{@{}lr@{}}
\toprule
Turns & 681 \\
Evictions (total) & 680 \\
Garbage collected & 74 \\
Read evictions (pageable) & 606 \\
Page faults & 659 \\
\textbf{Fault rate (total evictions)} & \textbf{97\% (659/680)} \\
Peak compression & 5{,}038KB $\to$ 339KB \\
Session termination & API rate limit \\
\bottomrule
\end{tabular}
\end{table}

The 97\% fault rate is a pathology, not a feature. The system
evicted almost everything and the model re-requested almost
everything. Three specific patterns emerged:

\emph{Thrashing cycle.} Three files (3{,}438 + 1{,}570 + 1{,}592
bytes) were evicted and faulted in a cycle across turns 163--170+.
The working set exceeded the resident set---classic
thrashing~\citep{denning1968working}.

\emph{Sequential scan.} During a planning phase, the model read
across three code repositories, cycling through the same 7~files
repeatedly. The working set for planning was larger than what the
age threshold allowed to remain resident.

\emph{Self-inflicted inflation.} The 5{,}038KB pre-compaction size
includes re-reads caused by previous evictions. Without the
eviction-fault cycle, the original content would have been
substantially smaller. The proxy was measuring its own overhead as
``bytes saved''---a metric artifact of thrashing.

\emph{Cost.} The session terminated by hitting the API rate limit,
not the context limit. At 659~faults (each an inference-priced tool
call round-trip), thrashing consumed rate budget faster than useful
work. This demonstrates that fault cost is not merely computational
but monetary.

\paragraph{Anchor handle behavior.}
When a fresh model instance resumed a session containing paged-out
content, it stated: ``Let me re-read the files I need since some
were paged out, then create the task list and start implementing.''
The model recognized the retrieval handles, understood what was
missing, and chose to fault content in before acting---without any
instruction to do so. The handle format is self-describing: the
model infers the recovery mechanism from the text of the summary
alone.


\section{Discussion}
\label{sec:discussion}

\subsection{Why This Waste Exists}

The append-only pattern in agentic AI tools is not an implementation
bug. It arises from four reinforcing factors:

\begin{enumerate}[nosep]
  \item \textbf{Training data}: Models learn from conversations that
    grow monotonically. No training examples demonstrate intelligent
    content removal.
  \item \textbf{API design}: The Messages API accepts a list of
    messages. The natural operation on a list is append. The API
    provides no mechanism for ``this content is stale.''
  \item \textbf{Framework defaults}: Every orchestration framework
    appends by default. Eviction requires explicit engineering.
  \item \textbf{Invisible cost}: Token consumption is billed after
    the fact. Quality degradation from context bloat is invisible
    unless measured. There is no backpressure signal.
\end{enumerate}

\subsection{The Inverted Cost Model}
\label{sec:inverted-cost}

In traditional virtual memory, keeping a page in physical memory is
free. RAM is allocated whether the page is accessed or not; the only
cost is \emph{faulting}---the latency of disk I/O when a needed page
is not resident. The entire replacement algorithm literature, from
FIFO to LRU to Clock to Working Set, optimizes a single objective:
\emph{minimize faults}.

In LLM context management, the cost model is inverted. Every token
kept in context costs money on every turn. A 5{,}000-token file
sitting in context for 20~turns costs 100{,}000 input tokens of
processing. Re-reading that file once when actually needed costs
5{,}000 tokens. The eviction saves 95{,}000 tokens. Keeping is
expensive; faulting is cheap.

This inversion changes the optimization objective. The correct
formulation is not ``minimize faults'' but:
\[
  \min \sum_{p \in P} \bigl[ C_{\text{keep}}(p) + C_{\text{fault}}(p) \bigr]
\]
where $C_{\text{keep}}(p) = |p| \cdot T_{\text{resident}}(p) \cdot
c_{\text{token}}$ is the cumulative cost of keeping page $p$ resident
for $T_{\text{resident}}$ turns, and $C_{\text{fault}}(p) = |p| \cdot
c_{\text{token}}$ is the one-time cost of restoring the page (one
turn of reprocessing). The break-even condition is:
\[
  |p| \cdot T_{\text{until\_next\_ref}}(p) \cdot c_{\text{token}} >
  |p| \cdot c_{\text{token}}
\]
which simplifies to: \emph{evict whenever the page will not be
referenced for more than one turn}. This is why FIFO works so well
in our system despite being the worst-performing policy in classical
VM: when keeping is expensive and faulting is cheap, aggressive
eviction is correct by default. Sophistication in replacement policy
is needed only to avoid evicting pages that will be needed on the
\emph{very next} turn.

\paragraph{Belady's MIN under inverted costs.}
Belady's optimal algorithm~\citep{belady1966study} evicts the page
whose next reference is farthest in the future, minimizing total
faults. Under the inverted cost model, the optimal policy is
different: it must minimize the sum of keeping costs and fault costs.
A page with a distant next reference should be evicted immediately
(saving many turns of keep cost at the price of one fault). A page
referenced every turn should never be evicted (the fault cost equals
one turn of keep cost, so eviction saves nothing). The optimal
offline policy under inverted costs is not MIN---it is a
cost-weighted variant where pages are evicted when their projected
keep cost exceeds their fault cost, regardless of whether a fault
will occur.

This has a practical implication: the \emph{size} of the page matters
for eviction priority in a way it does not in classical VM (where all
pages are equal size). A 10{,}000-token file costs ten times as much
per turn as a 1{,}000-token file. Large pages should be evicted
eagerly unless access frequency justifies the keep cost.

\paragraph{Non-linear fault cost.}
The linear fault cost $C_{\text{fault}}(p) = |p| \cdot c_{\text{token}}$
above is a simplification. Transformer inference is dominated by
self-attention, which is $O(n^2)$ in sequence length. A page fault
requires an additional full inference pass---the model emits a
\texttt{tool\_use} for the recall handle, and the restored content
returns as a \texttt{tool\_result} message that triggers a second
inference over the entire context. The true fault cost is therefore
proportional to $n^2$ at the current context size $n$, not to the
page size $|p|$ alone.

This produces a counter-intuitive policy gradient. At low fill
(e.g., $n = 40\text{K}$), faults are cheap and aggressive eviction is
correct---the quadratic cost of an extra pass is modest, and keeping
the working set small reduces the $n^2$ base for all subsequent turns.
At high fill (e.g., $n > 100\text{K}$), faults become expensive: the
additional inference pass costs roughly $(n + |p|)^2 \approx n^2$
tokens of compute. The eviction policy should therefore become
\emph{more conservative} as context pressure rises---the opposite of
the naive instinct to evict aggressively under pressure. At high
fill, the system should evict only content it is highly confident will
not be referenced again.

\paragraph{Object-based cost variance.}
Unlike classical VM, where all pages are fixed-size blocks (typically
4\,KB), Pichay manages variable-size objects: a recalled tensor may be
a 3-line summary or a 200-line file. This means eviction and fault
costs vary by orders of magnitude across the working set. The
replacement policy cannot treat all eviction candidates as equivalent;
it must weight both the per-turn keep cost and the potential fault cost
by object size. Combined with the quadratic fault penalty, this argues
for a \emph{size-aware, fill-sensitive} replacement policy that has no
direct analogue in the classical VM literature.

\paragraph{Cache invalidation cost.}
The cost model so far considers only token-level costs (keeping and
faulting). Structural mutations---collapse operations that remove or
replace blocks in the message array---have an additional cost:
\emph{prompt cache invalidation}. Inference providers cache the
tokenized prefix of repeated requests; when the prefix changes, the
cache misses and the entire context must be reprocessed.

In production, a collapse operation that compressed 12~turns of
orientation dialogue into a single summary sentence caused the cache
hit rate to drop from 100\% to 25\% for one turn, then recover to
100\% on the following turn as the new prefix stabilized. The cost
was one full recompute of $\sim$105K tokens---comparable to the
cost of several page faults. A collapse that saves 10KB of context
but invalidates a 100K-token cached prefix is a net loss unless the
space savings persist for enough subsequent turns to amortize the
one-time recompute.

This argues for \emph{batching} structural mutations: accumulate
collapse candidates and execute them together in a single pass,
paying the cache invalidation cost once rather than per-operation.
It also argues against frequent small collapses in favor of
infrequent large ones.

\paragraph{Pin decay.}
The inverted cost model also argues against permanent pins. In our
current design (Section~\ref{sec:pinning}), one fault pins a page
permanently. But a fault tells us the content was needed \emph{then},
not forever. Under the cost model, a pin should decay: its strength
halves every $K$~turns since last access, and the page becomes
evictable when the projected keep cost of the remaining pin lifetime
exceeds the fault cost. This gives LRU-like behavior with
cost-weighted decay, and prevents the monotonic working-set growth
that permanent pinning causes in long sessions.

\subsection{Generalizability}

The waste patterns we measure are structural consequences of the
agentic architecture:

\begin{itemize}[nosep]
  \item Any system that sends tool definitions on every request
    will send schemas for unused tools. The specific percentage
    depends on tool count and usage distribution, but the mechanism
    is universal.
  \item Any system that injects instructions into messages risks
    duplication. Claude Code's triplication is egregious but not
    unique.
  \item Any system that keeps tool results in context for the
    session lifetime will see amplification proportional to session
    length.
\end{itemize}

Our empirical evidence is from one system (Claude Code) and one
user. The structural argument is strong---the architecture
forces the waste---but comparative measurement of other tools
would strengthen the generalizability claim.

\subsection{Scope, Non-Goals, and Threats to Validity}

\paragraph{Scope.}
This paper evaluates context waste and demand paging behavior for
agentic coding workloads with persistent tool and transcript history.
The target claim is mechanism-level: when requests repeatedly resend
large static prefixes and stale tool output, eviction plus fault-based
restoration reduces cumulative input processing.

\paragraph{Non-goals.}
We do not claim universal percentage reductions across all assistants,
all model families, or all task domains. We do not evaluate frontier
reasoning benchmarks, autonomous planning quality, or human preference
at product scale. We also do not claim that L1 expansion (larger
context windows) is obsolete; we claim it is incomplete without
hierarchical management.

\paragraph{Threats to validity.}
External validity is limited by the single-user corpus and one primary
assistant implementation. Internal validity risks include prompt drift,
toolset changes across versions, and workload phase effects that alter
hot-set composition. To mitigate drift, we report a frozen cohort for
headline numbers and provide executable recount tooling plus
timestamped live snapshot artifacts for auditability.

\subsection{Quality: LLM-Judged Equivalence Check}
\label{sec:quality}

The intervention must not degrade output quality. We therefore run a
paired equivalence-style check: if results are broadly comparable while
compute drops, the intervention is useful.

\paragraph{Protocol.}
We select 18~sessions from the corpus (426--1{,}354~messages,
340K--718K~characters of context). For each session, we construct
paired contexts at 65--75\% of the conversation: the \emph{baseline}
retains all messages; the \emph{treatment} replaces consumed tool
results outside a 20-message recency window with tombstones, modeling
\textsc{Pichay}'s eviction behavior. Both conditions receive the same
continuation prompt (the next user message). Outputs are generated by
Sonnet~4 under both conditions.

An ensemble of three LLM judges (one Sonnet~4, two Haiku~4.5)
evaluates each paired output on correctness, completeness, and
coherence (1--5~scale). Judges also state a blind preference
(A, B, or tie) and whether they can identify which output had
reduced context. A/B assignment is randomized per judge to
prevent position bias; scores are remapped to baseline/treatment
before aggregation.

\paragraph{Results.}

\begin{table}[ht]
\centering
\caption{Non-inferiority evaluation: 18~sessions, 54~verdicts.
  Treatment evicts consumed tool results outside a 20-message
  recency window (mean compression 48\%).}
\label{tab:noninferiority}
\small
\begin{tabular}{@{}lrr@{}}
\toprule
\textbf{Metric} & \textbf{Baseline} & \textbf{Treatment} \\
\midrule
Judges preferring & 15 (28\%) & 20 (37\%) \\
Ties & \multicolumn{2}{c}{19 (35\%)} \\
\midrule
Mean correctness & 3.89 & 3.74 \\
Mean completeness & 3.59 & 3.59 \\
Mean coherence & 3.74 & 3.69 \\
Max score $\Delta$ & \multicolumn{2}{c}{0.15} \\
\midrule
Detection rate & \multicolumn{2}{c}{57\% ($p = 0.14$, n.s.)} \\
\bottomrule
\end{tabular}
\end{table}

Judges prefer the treatment \emph{more often} than the baseline
(37\% vs.\ 28\%), with 35\% ties. Completeness scores are identical;
correctness and coherence differ by at most 0.15~points on a 5-point
scale. Detection rate (57\%) is not significantly above chance
(binomial $z = 1.09$, $p = 0.14$, $n = 54$).

\paragraph{Failure modes.}
Two sessions (11\%) produced degenerate treatment outputs (zero
characters): the continuation prompt implicitly referenced content
that existed only in a tombstoned tool result. These are genuine
eviction casualties. The failure pattern is specific: the user's
prompt references a recently-consumed tool result by content rather
than by name. A reference-aware eviction policy would detect such
forward dependencies; the current heuristic (consumed + outside
recency window = evictable) does not.

\paragraph{The attention-concentration effect.}
In several sessions, all three judges preferred the treatment output.
Removing half the context did not merely preserve quality---it
improved it. This is consistent with the attention-dilution hypothesis:
transformer attention distributes weight across all tokens in context;
irrelevant tokens (stale tool results, consumed intermediate outputs)
dilute attention on the tokens that matter. Evicting them concentrates
attention on signal. The effect is strongest in sessions with high
tool-result density, where the evicted content is genuinely noise.
We note this as preliminary evidence; establishing the effect
rigorously would require controlled attention analysis with larger~$N$.

\subsection{Compute and Energy Implications}
\label{sec:energy}

The 970~million tokens of addressable waste in our corpus have
three compute cost dimensions:

\paragraph{Input processing.}
At a 93.5\% cache hit rate, most saved tokens are cache reads
(907M tokens at reduced cost). However, 63M tokens are cache
creation or uncached---full forward-pass computation that need not
occur.

\paragraph{Attention cost.}
For every output token, the model computes attention over the full
context. Removing 17{,}913 tokens from a mean context of 82{,}061
reduces the per-output-token attention computation by 21.8\%.
Across the corpus, this eliminates 85~billion token-token attention
pairs.

\paragraph{Throughput.}
Each token in context requires KV cache memory. For large models,
the per-token KV cache footprint is on the order of megabytes.
Smaller context per request means less KV cache memory per request,
which means more concurrent requests per GPU. A 21.8\% context
reduction translates, to first approximation, to 21.8\% more
concurrent users on the same GPU fleet.

This is the argument that changes the infrastructure economics.
The savings are not just ``less cost per request''---they are
``more requests per machine,'' which is the difference between
building new data centers and not building them.

\paragraph{Compounding across sessions.}
These per-call savings compound. Waste prevented on turn~$N$ is not
merely saved once---it is absent from every subsequent turn's
context, where it would have been reprocessed under $O(n^2)$
attention cost. The total compute savings across a session are
therefore superlinear in the waste fraction and grow with session
length. For a 100-turn session with 21.8\% addressable waste, the
cumulative attention savings are not 21.8\% of total session
compute---they are substantially larger, because the quadratic cost
of carrying unnecessary tokens compounds on every subsequent call.
The infrastructure implication follows directly: the fastest tokens
are the ones you never process.

\begin{figure}[t]
\centering
\begin{tikzpicture}
\begin{axis}[
    width=\columnwidth,
    height=0.6\columnwidth,
    xlabel={Turn},
    ylabel={Cumulative input tokens (millions)},
    xmin=0, xmax=88,
    ymin=0, ymax=9.5,
    legend pos=north west,
    legend style={font=\small},
    grid=major,
    grid style={gray!20},
    y tick label style={/pgf/number format/fixed,
                        /pgf/number format/precision=1},
]

\addplot[thick, red!70!black, mark=none, name path=baseline] coordinates {
    (0, 0.013651) (10, 0.352856) (20, 1.037825) (30, 1.850237)
    (40, 2.884656) (50, 3.989325) (60, 5.146048) (70, 6.379942)
    (80, 7.689415) (87, 8.645027)
};
\addlegendentry{Baseline (full context)}

\addplot[thick, blue!70!black, mark=none, dashed, name path=managed] coordinates {
    (0, 0.007508) (10, 0.194065) (20, 0.570793) (30, 1.017614)
    (40, 1.586539) (50, 2.194101) (60, 2.830295) (70, 3.508933)
    (80, 4.229138) (87, 4.754720)
};
\addlegendentry{Managed (Pichay)}

\addplot[red!10, opacity=0.5, forget plot]
    fill between[of=baseline and managed];

\end{axis}
\end{tikzpicture}
\caption{Cumulative input tokens processed over an 88-turn session.
  Without context management, every turn re-processes all prior
  content, producing superlinear cumulative cost. Pichay's eviction
  reduces cumulative token spend by 45\% (\$11.67 at Opus pricing).
  The shaded area represents wasted computation.}
\label{fig:cumulative-cost}
\end{figure}
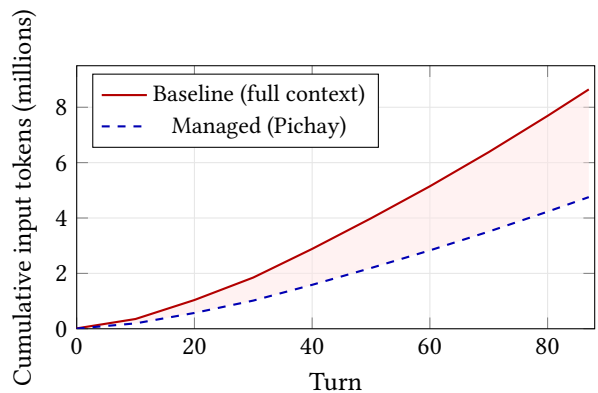

Figure~\ref{fig:cumulative-cost} illustrates this compounding on an
88-turn live session running through \textsc{Pichay}. Cumulative
baseline cost (processing all context every turn) reaches 8.6M~input
tokens; managed cost reaches 4.8M---a 45\% cumulative reduction,
substantially larger than the per-turn compression ratio, because
each evicted token is absent from \emph{every subsequent turn}.

\paragraph{Fleet-scale extrapolation.}
Our corpus represents one user over four months. A fleet-wide
projection requires per-user token consumption data that inference
providers hold but have not published. The extrapolation framework
is straightforward: $N$ users $\times$ 970M tokens per user $=$
total addressable waste. The mechanism (tool schema waste, content
duplication, result amplification) is architectural, so the
per-user waste fraction should be relatively stable even if
absolute token counts vary.


\section{Future Work}
\label{sec:future}

\paragraph{L3 evaluation.}
The collapse mechanism (Section~\ref{sec:l3}) is implemented but not
yet evaluated at scale. Open questions include: how much context does
model-authored compaction recover compared to system-initiated
eviction? Do declared losses in collapse summaries enable accurate
fault decisions (can the model determine from a summary alone whether
it needs the original content)? What is the quality impact of
replacing 20~turns of dialogue scaffolding with a single outcome
sentence? These require longitudinal measurement across sessions of
varying length and task complexity.

\paragraph{Cost-aware eviction pressure.}
The current eviction policy uses a space-based threshold (context
window capacity). The inverted cost model argues for cost-based
pressure: 23\% memory pressure sounds low, but 45{,}000 tokens per
turn at inference pricing is real money. Eviction pressure should be
denominated in token-turns (the product of page size and projected
remaining lifetime), not in fractional capacity. This reframes
eviction from ``are we running out of space?'' to ``is this page
earning its keep?''

\paragraph{Cost-weighted pin decay.}
Pinned pages should not remain resident forever. A fault tells us
the content was needed \emph{then}, not forever. We propose
cost-weighted decay: pin strength halves every $K$~turns since last
access. A page becomes evictable when its projected keep cost exceeds
its fault cost. Under low memory pressure, this is permissive
(pins persist). Under high pressure, cold pins are released first.
This gives LRU-like behavior with cost weighting, and prevents the
monotonic working-set growth that permanent pinning causes in long
sessions.

\paragraph{Phase-aware eviction.}
Planning and execution have different working set characteristics.
Planning requires broad simultaneous context (many files held at
once); execution is sequential (read, edit, move on). The proxy
could detect phase from request patterns---many Reads with no Edits
suggests planning---and adjust the eviction threshold accordingly.

\paragraph{Per-connection isolation.}
The current implementation uses a single PageStore for all
connections. Subagent sessions share eviction state with the main
session, causing cross-contamination. Per-connection isolation would
give each session its own eviction history, fault tracking, and
pinning state.

\paragraph{Trace-driven simulation.}
We extract reference strings (the sequence of page accesses) from
session transcripts and replay them under different replacement
policies. This enables evaluation of policies against recorded
workloads without live inference-priced experiments. The inverted cost
model (Section~\ref{sec:inverted-cost}) means the optimal offline
policy is not Belady's MIN but a cost-weighted variant that minimizes
total keep cost plus fault cost. We plan to derive and evaluate this
cost-optimal offline bound alongside MIN for comparison.

\paragraph{Cross-session access pattern prediction.}
The break-even formula requires estimating $T_{\text{until\_next\_ref}}$---the
number of turns until a page is next referenced. Within a single
session, this can only be estimated from local patterns. Across
sessions, the prediction becomes tractable: coding sessions exhibit
regular access patterns (orientation reads of project structure,
iterative edits of a working file, periodic reference to configuration).
A Markov model trained on cross-session reference strings could
predict re-reference probability per page per turn, enabling
cost-optimal eviction decisions. The data collection is nearly free:
the proxy already logs every eviction, every fault, every tool call.
What is needed is a persistent, queryable store of access patterns
that survives session boundaries---precisely the role of a cross-session
memory system.

\paragraph{Cooperative extensions.}
Phantom tools open a side-channel between the model and the proxy.
The current vocabulary---release and fault---is minimal. Extensions
include: \emph{prefetch hints} (model signals upcoming working set
needs), \emph{eviction classes} (groups of pages evicted as a unit),
\emph{variable-fidelity tombstones} (AST skeletons instead of bare
path markers), and \emph{ephemeral flags} (tool results marked for
immediate eviction after extraction). Each expands the model's
ability to communicate cognitive state to the memory manager.

\paragraph{Beyond paging: object-addressed memory.}
The current system manages fixed-granularity pages addressed by file
path. But the natural unit for LLM memory management is the semantic
object---a conversation phase, a design decision, a debugging session.
These objects vary in size, have relationships to each other, and can
be compressed to multiple fidelity levels rather than the binary
resident/evicted state of hardware pages. The eviction tombstone
becomes a compressed summary that functions as both a page table
entry (retrieval handle to the backing store) and a cache line
(enough semantic content to answer queries without faulting).
Declared losses in the summary tell the model what it \emph{cannot}
answer from the summary alone, guiding the decision to fault.
The backing store is queryable---the model can ask a question of
evicted content without materializing it in full. This shifts the
abstraction from block-addressed paging to object-addressed memory
management, where eviction is cooperative, compression is authored,
and the backing store is a database rather than a swap partition.

\paragraph{Comparative measurement.}
Instrument other agentic tools (Cursor, Copilot, Windsurf,
Continue.dev) to measure the same metrics. The structural argument
predicts similar waste profiles; empirical confirmation would
strengthen generalizability.


\section{Conclusion}

The context window of a large language model is not memory. It is
L1~cache---a small, fast, expensive level in what should be a
managed hierarchy. The field treats it as the entire memory system.
There is no L2, no paging, no eviction policy, no working set
estimation. The result: 21.8\% structural waste in 4.45~billion
measured tokens, with the model reprocessing stale content at
84.4$\times$ amplification.

\textsc{Pichay} demonstrates that the virtual memory abstractions
developed for physical memory in the 1960s apply directly to LLM
context windows. Demand paging with a simple FIFO policy recovers
36~percentage points of context capacity in steady-state use.
Fault-driven pinning reduces repeated faults on working-set content in
steady-state use.
Retrieval handles---designed as space-saving markers---function as
late-binding anchors that models understand without instruction.
Graduated pressure zones and cooperative cleanup tags give the model
agency over its own memory management---a capability with no analogue
in hardware memory hierarchies.

The system requires no changes to models, clients, or inference
APIs. It is implemented as a transparent proxy and deployed in
production use. The paper you are reading was written through it.

The deeper contribution is the architectural observation: LLM
systems need not one large context window but a hierarchy of
managed levels---from the generation window (L1) through persistent
cross-session memory (L4)---connected by eviction policies and
fault mechanisms. The first three levels are deployed; the first
two are evaluated. L3---model-initiated conversation compaction---is
the newest mechanism, and its evaluation is ongoing. The solutions
for the remaining level exist in the OS literature.
What remains is to connect them.


\section*{Authorship and AI Contribution}
\label{sec:authorship}

This paper was co-authored by a human systems researcher and an AI
system (Claude Opus, Anthropic). The human contributed the memory
hierarchy insight, research direction, experimental design, and
editorial judgment---including the observation that context
management is memory management, which reframed the entire project.
The AI contributed code (analysis tools, proxy implementation,
paging system), data analysis, and paper drafting.

\textsc{Pichay} was built by the system being studied---Claude Code
running through its own proxy---and the paper was written through it.
The evaluation data in Section~\ref{sec:production} was generated as
a side effect of the system's own development. We consider this
self-referentiality a feature: the system's ability to measure,
optimize, and sustain its own context usage over hundreds of turns is
the strongest evidence that the interventions are practical.

All instruments and data are open source at:
\url{https://github.com/fsgeek/pichay}.
The paper snapshot cited in this manuscript is release
\texttt{v0.1.0-paper} at commit \texttt{b56701a}.
An immutable archival snapshot is available via Zenodo DOI:
\url{https://doi.org/10.5281/zenodo.18930122}.


\appendix
\section{Data Snapshot and Counting Protocol}
\label{sec:appendix-protocol}

To keep counts reproducible as the live corpora evolve, we separate
paper numbers (frozen cohort for this draft) from live recounts.
The counting protocol is versioned in the repository:
\texttt{docs/corpus\_protocol.md}.

\paragraph{Counting tool.}
Session counting is performed by
\texttt{tools/corpus\_counts.py}, which classifies Claude Code JSONL
files into \texttt{main}, \texttt{subagent}, \texttt{compact},
\texttt{prompt\_suggestion}, and \texttt{other}, and reports both
raw and content-hash-deduplicated totals.

\paragraph{Recount command.}
\begin{verbatim}
./tools/reproduce_paper_counts.sh
\end{verbatim}
This writes a timestamped snapshot JSON to
\texttt{paper/data/corpus\_snapshot\_YYYYMMDD\_live.json}.

\paragraph{5-minute reproduction path.}
From repository root:
\begin{verbatim}
./tools/reproduce_paper_counts.sh
python3 tools/check_paper_numbers.py \
  --paper paper/main.tex \
  --snapshot paper/data/corpus_snapshot_20260307_live.json
latexmk -pdf paper/main.tex
\end{verbatim}
Expected outputs: (1) a new snapshot JSON under \texttt{paper/data/},
(2) a paper-number consistency report, and (3) \texttt{paper/main.pdf}.
On a typical laptop this path completes in minutes; the recount step is
I/O-bound on local corpus size.

\paragraph{Current live recount artifact (2026-03-07).}
\texttt{paper/data/corpus\_snapshot\_20260307\_live.json}
with roots:
\texttt{\~/.claude/projects} and
\texttt{\~/projects/yanantin/tmp/ubuntu-vm.claude/projects},
size filter \texttt{min\_size=10000}, and Claude Desktop summary
from \texttt{\~/projects/yanantin/tmp/claude-desktop/conversations.json}.

For this paper version, headline results use a frozen cohort to
avoid drift across revisions; live recount artifacts are provided for
auditability.


\bibliographystyle{ACM-Reference-Format}
\bibliography{references}

\end{document}